\documentclass[prl,aps,amssymb,superscriptaddress,twocolumn]{revtex4}
\usepackage{graphicx}
\usepackage{amsmath}
\usepackage{mathtools}
\usepackage{textcomp}

\begin{document}

\title{Observation of Squeezing in the Electron Quantum Shot Noise of a Tunnel Junction}

\author{Gabriel Gasse}
%\affiliation{Laboratoire de Physique des Solides, CNRS UMR8502, Univ. Paris-Sud 11, F91405 Orsay, France}
\affiliation{D\'{e}partement de Physique, Universit\'{e} de Sherbrooke, Sherbrooke, Qu\'{e}bec, Canada, J1K 2R1}

\author{Christian Lupien}
\affiliation{D\'{e}partement de Physique, Universit\'{e} de Sherbrooke, Sherbrooke, Qu\'{e}bec, Canada, J1K 2R1}

\author{Bertrand Reulet}
\affiliation{D\'{e}partement de Physique, Universit\'{e} de Sherbrooke, Sherbrooke, Qu\'{e}bec, Canada, J1K 2R1}

\date{\today}
\begin{abstract}
We report the measurement of the fluctuations of the two quadratures of the electromagnetic field generated by a quantum conductor, a dc- and ac-biased tunnel junction placed at very low temperature. We observe that the variance of the fluctuations on one quadrature can go below that of vacuum, i.e. that the radiated field is squeezed. This demonstrates the quantum nature of the radiated electromagnetic field.
\end{abstract}
\maketitle

A great effort is currently deployed to find sources of quantum light. A light with properties beyond that of classical physics is indeed essential in the development of quantum information technology \cite{Braunstein,Weedbrook} and has direct applications in metrology \cite{Caves}. Quantum light can be non-classical in several ways. Squeezed light offers the possibility to beat vacuum fluctuations along one quadrature: the rms flutuations $\Delta A^2$ of the amplitude of $A\cos(\omega t)$ can be smaller than that of vacuum, at the expense of an increase of the rms fluctuations $\Delta B^2$ of the (quadrature) amplitude of $B\sin(\omega t)$, in order to preserve the Heisenberg uncertainty principle (for a review on squeezing, see \cite{Loudon,Gardiner,Wiseman}).

Many systems have been invented to produce squeezed light, based for example on non-linear crystals, atomic transitions and non-linear cavities in optics \cite{Slusher} but also with parametric amplifier and qubits in the microwave domain \cite{Yurke,Movshovich,Nation,Devoret}. The key ingredient in all these systems is the existence of a nonlinearity, which allows the mixing of vacuum fluctuations with the classical, large field of a coherent pump. Here we use the discreteness of the electron charge $e$ as a source of non-linearity. A tunnel junction (two metallic contacts separated by a thin insulating layer) has a \emph{linear} $I(V)$ characteristics at low voltage and thus cannot be used as a non-linear element to mix signals. There is no photo-assisted dc transport, i.e. no rectification. However, it exhibits shot noise, i.e. the variance of the current fluctuations $\Delta I^2$ generated by the junction depends on the bias voltage. For example, at low frequency and high current, the noise spectral density is given by $e|I|$ (for a review on shot noise in mesoscopic conductors, see \cite{BuBlan,Nazarov_book}). Thus, in the presence of an ac excitation, the junction exhibits photo-assisted noise \cite{Lesovik,Rob,Kozhevnikov} as well as a dynamical modulation of its noise \cite{GR1,GR2}. We use this modulation the \emph{intrinsic} noise of the junction by an external ac excitation to generate squeezing.

In this letter we report the measurement of the variance of the two quadratures of the electromagnetic field around frequency $\omega/2\pi=7.2$ GHz generated by a tunnel junction placed at ultra-low temperature and biased by both a dc voltage $V_{dc}$ and an ac voltage $V_{ac}$ at frequency $\omega_0=2\omega$ or $\omega_0=\omega$. We show that depending on $V_{dc}$ and $V_{ac}$, the electromagnetic field generated by the junction can be squeezed, in very good agreement with theoretical predictions.

\begin{figure}[montage]
\includegraphics[width=0.9\linewidth]{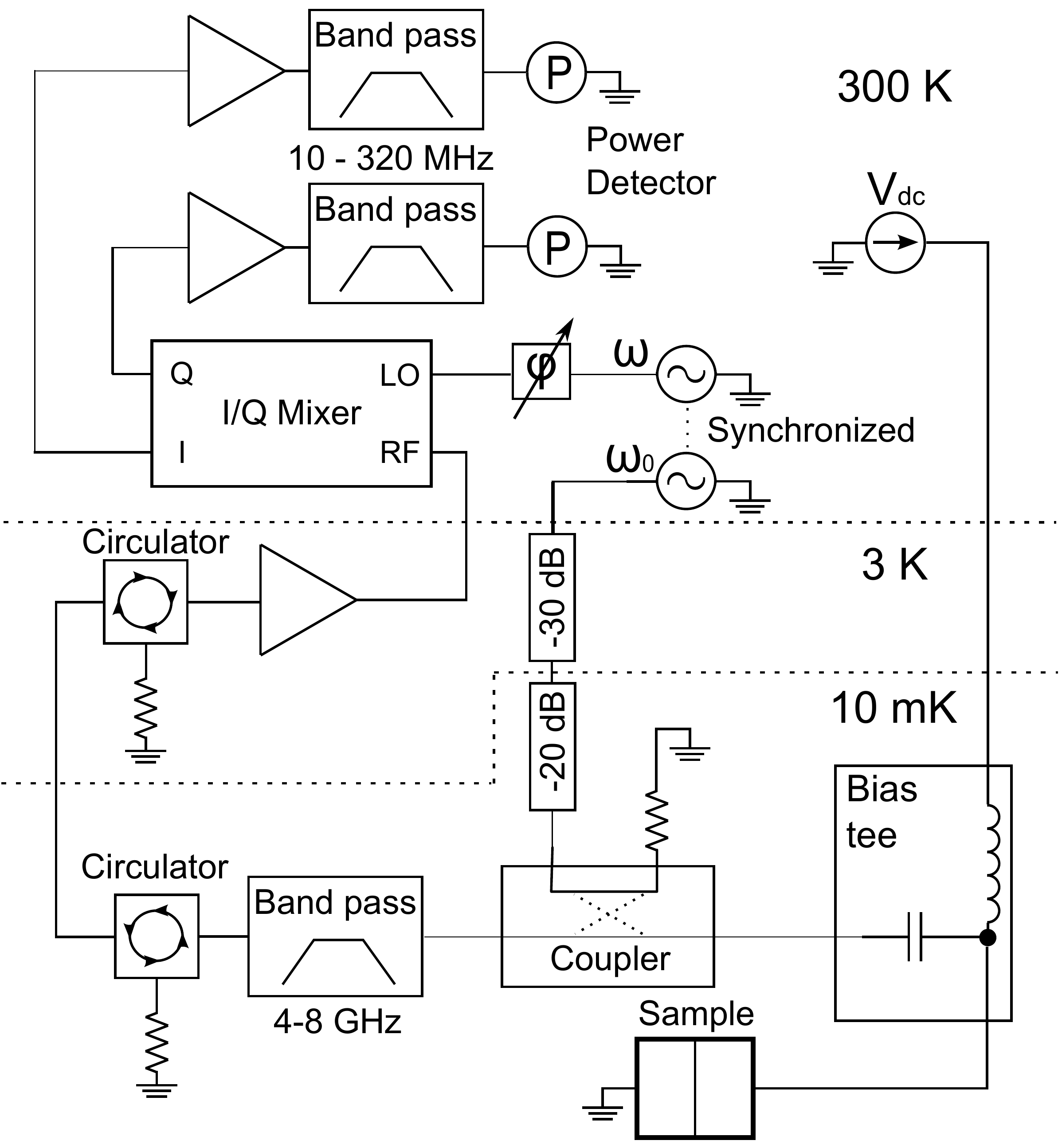}
\caption{Experimental setup. The \textcircledP \: symbol represents a power detector.}
\label{fig:setup}
\end{figure}

\emph{Experimental setup.}
We cooled to 10 mK in a dilution refrigerator an Al/ Al oxide/ Al tunnel junction of resistance $R = $70$\Omega$, similar to that used for noise thermometry \cite{Lafe}. We applied a 50~mT perpendicular magnetic field to keep the aluminum normal even at the lowest temperatures. The detection setup is depicted on Fig. \ref{fig:setup}. The tunnel junction is dc biased through a bias tee while the ac current fluctuations it generates are bandpass filtered in the range 4-8 GHz and amplified before exiting the cryostat. The presence of cryogenic circulators between the sample and the amplifier helps keep the electron temperature low while a coupler allows to excite the junction at high frequency (after strong attenuation). The detected electromagnetic field is separated into two quadratures around frequency $\omega$ thanks to an IQ mixer, which provides $A(t)$ and $B(t)$ with a bandwidth of $500$MHz. The rms voltage of the two quadratures, i.e. $\Delta O^2=\langle [O(t)-\langle O(t)\rangle ]^2\rangle$ with $O=A,B$  are simultaneously measured by two power detectors. The excitation of the sample at frequency $\omega_0=2\omega/p$ with $p=1,2$ is performed by a second microwave generator synchronized with the one used as a reference for the IQ mixer. The amplitude and phase of the two generators can be tuned independently.

\begin{figure}
\includegraphics[width=1.1\columnwidth]{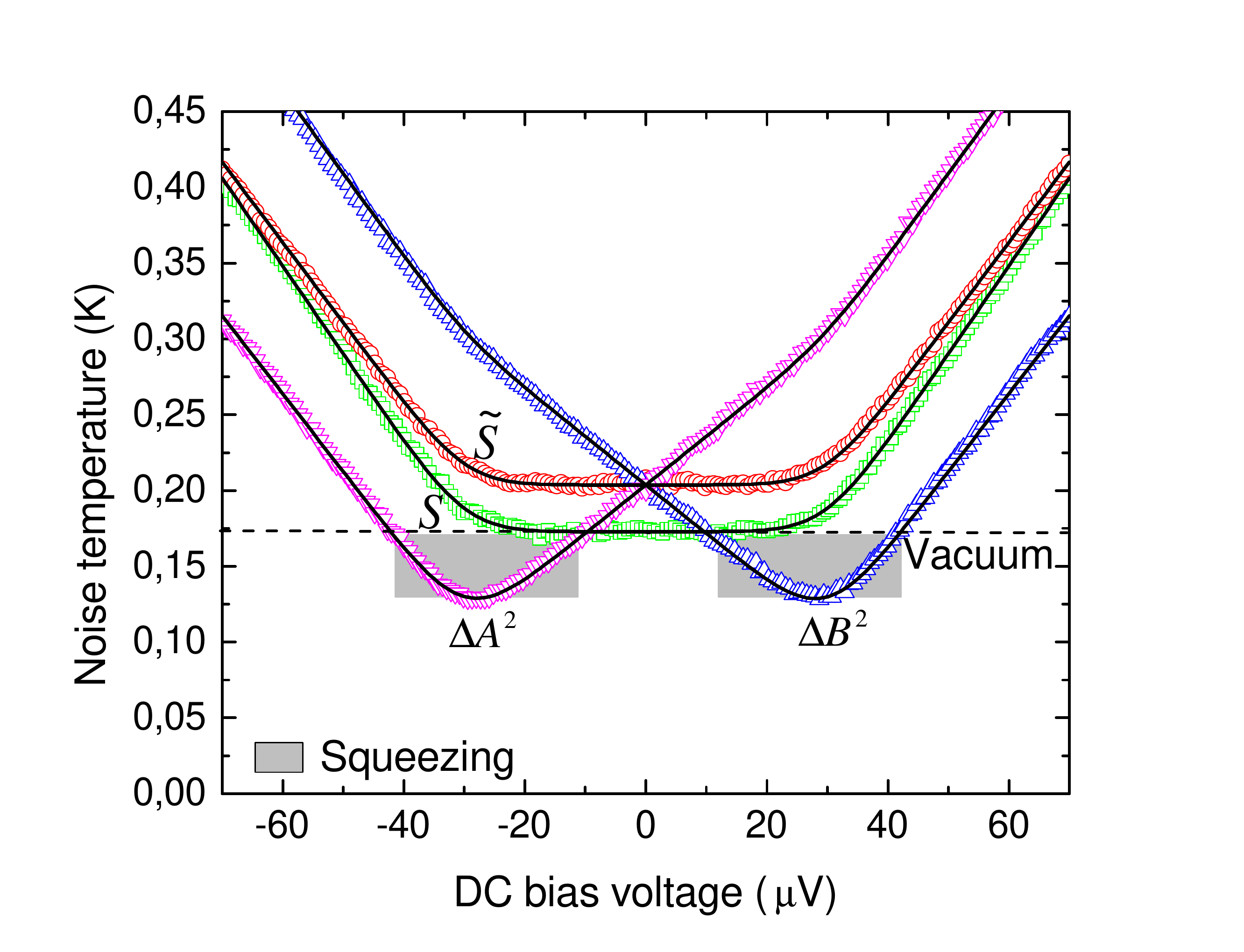}
\caption{Reduced $\Delta A^2$ or $\Delta B^2$ for $\omega_0=2\omega$. Green squares (label "S"): $V_{ac}=0$ so $\Delta A^2 = \Delta B^2=S$). Orange squares (label "$\widetilde{S}$"): $V_{ac}=46\mu V$ but generators slightly detuned, so $\Delta A^2=\Delta B^2=\widetilde{S}$. Purple squares: $\Delta A^2$ for $V_{ac}=46\mu V$. Blue squares: $\Delta B^2$ for $V_{ac}=46\mu V$. The dotted line corresponds to vacuum fluctuations, $\hbar\omega/(2k_B)$. The shaded region (below vacuum fluctuations) indicates squeezing.}
\label{fig:w0=2w}
\end{figure}

\begin{figure}
\includegraphics[width=1.1\columnwidth]{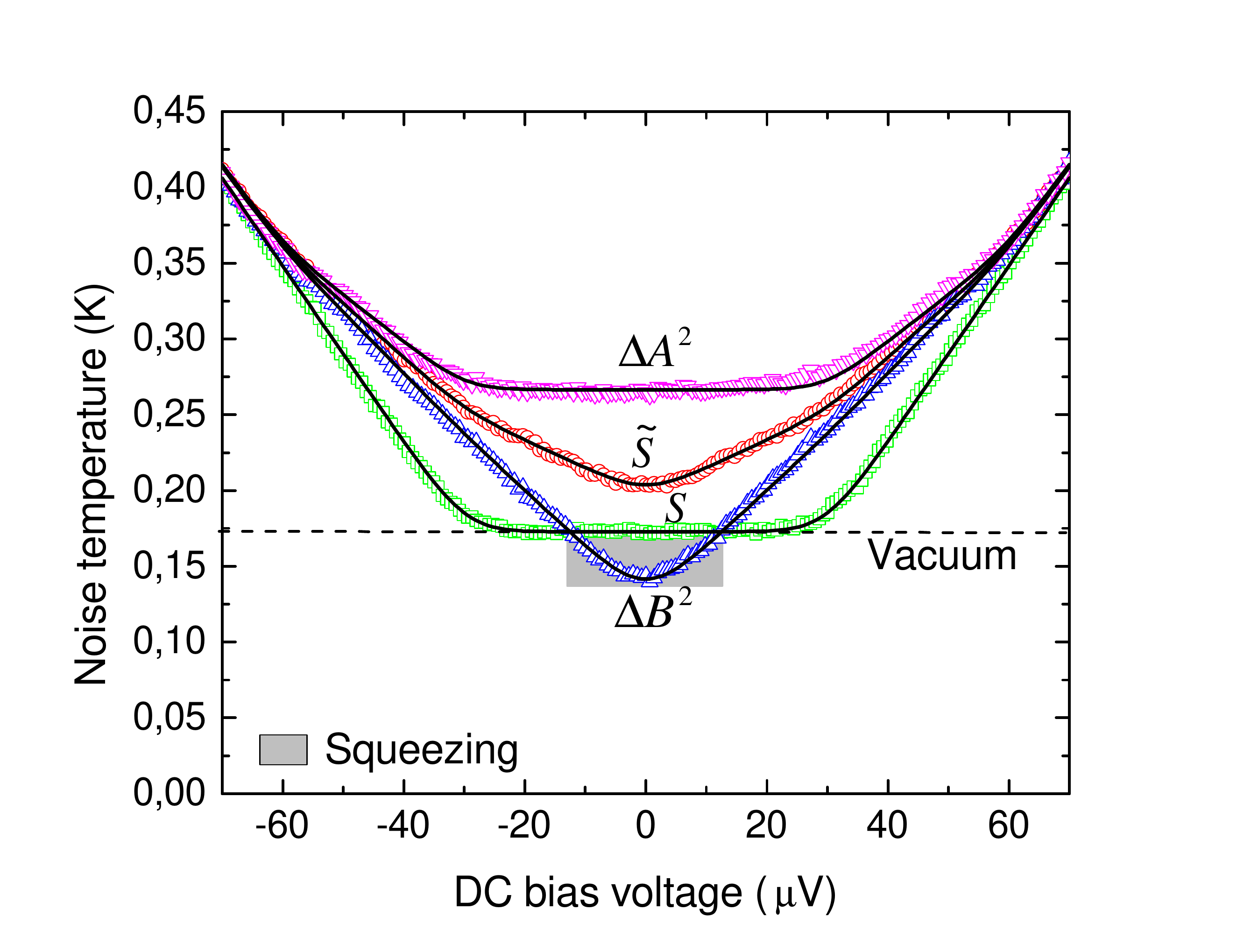}
\caption{Reduced $\Delta A^2$ or $\Delta B^2$ for $\omega_0=\omega$. Green squares (label "S"): $V_{ac}=0$ so $\Delta A^2 = \Delta B^2=S$). Orange squares (label "$\widetilde{S}$"): $V_{ac}=36\mu V$ but generators slightly detuned, so $\Delta A^2=\Delta B^2=\widetilde{S}$. Purple squares: $\Delta A^2$ for $V_{ac}=36\mu V$. Blue squares: $\Delta B^2$ for $V_{ac}=36\mu V$. The dotted line corresponds to vacuum fluctuations, $\hbar\omega/(2k_B)$. The shaded region (below vacuum fluctuations) indicates squeezing.}
\label{fig:w0=w}
\end{figure}

\emph{Results.}
We first consider the noise measured in the absence of ac excitation of the junction. In this case nothing sets an absolute phase in the measurement, so $\Delta A^2=\Delta B^2$, as shown on Figs. \ref{fig:w0=2w} and \ref{fig:w0=w} (green squares). This measurement is equivalent to the usual measurement of the variance of current fluctuations of the sample with a power detector, i.e. $\Delta A^2=\Delta B^2 = G(S_{amp}+S(V_{dc},\omega))$ with $G$ the gain of the setup, $S_{amp}$ the current noise spectral density of the amplifier and $S(V_{dc},\omega)$ the noise spectral density of the current fluctuations in the sample at frequency $\omega$ for a bias voltage $V_{dc}$, given by $S(V_{dc},\omega)=[S_0(V_{dc}+\hbar\omega/e)+S_0(V_{dc}-\hbar\omega/e)]/2$ with $S_0(V)=R^{-1}eV\coth(eV/2k_BT)$ the zero frequency shot noise. From this measurement we extract $G$, $S_{amp}$ and the temperature of the electrons $T=28$mK. $S(V_{dc},\omega)$ is constant as long as $eV<\hbar\omega$, see the wide plateau around $V_{dc}=0$ on Figs. \ref{fig:w0=2w} and \ref{fig:w0=w}. In this regime no photon is emitted by the junction and the current fluctuations we observe correspond to the vacuum fluctuations: $S_{vac}=S(V_{dc}=0,\omega)=R^{-1}\hbar\omega$, i.e. an equivalent noise temperature $T_{vac}=\hbar \omega / (2 k_B)=180$mK. In the following, all the measurements will be given in terms of noise temperature $T_N=R\Delta^2/(2k_B)$ with $\Delta^2=\Delta A^2$ or $\Delta B^2$.

In the presence of an ac excitation at frequency $\omega_0$ and amplitude $V_{ac}$ the spectral density of current fluctuations $\widetilde{S}(V_{dc},\omega,V_{ac},\omega_0)$ is given by:
\begin{equation}
\widetilde{S}=\sum_{n=-\infty}^\infty J_n\left(\frac{eV_{ac}}{\hbar \omega_0}\right)
S\left(V_{dc}+n\frac{\hbar\omega_0}{e},\omega\right)
\label{eq:Sp}
\end{equation}%
This quantity, the so-called photo-assisted noise, has been first predicted in  \cite{Lesovik} and observed in \cite{Rob,Kozhevnikov}. From the measurement of $\widetilde{S}$, see Figs \ref{fig:w0=2w} and \ref{fig:w0=w} (orange squares), we can calibrate the excitation power, i.e. know $V_{ac}$ experienced by the sample. This measurement is obtained with our setup by detuning the two microwave generators by $100$kHz. This setting is equivalent to set $\omega=2\omega_0/p$ and average over the phase difference between the two generators. In that case one has
$\Delta A^2=\Delta B^2=\widetilde{S}$.

We now turn to the case where excitation and detection are synchronized, i.e. we perform phase sensitive noise measurements. An example of measurement of  $\Delta A^2$ and $\Delta B^2$ (which are now different) vs. dc voltage is shown on Fig.\ref{fig:w0=2w} for an excitation at frequency $\omega_0=2\omega$ and on Fig.\ref{fig:w0=w} for $\omega_0=\omega$. We observe that in both cases there is a range of $V_{dc}$ where one quadrature is below the plateau corresponding to vacuum fluctuations: this corresponds to squeezing of the electromagnetic field generated by the junction. We have performed such measurements for many excitation powers. The optimal squeezing for $\omega_0=2\omega$ (Fig. \ref{fig:w0=2w}) is found to occur at $V_{dc}\simeq \hbar\omega/e$ and corresponds to $0.74$ times vacuum fluctuations, i.e. $1.31$dB below vacuum. The optimal squeezing for $\omega_0=\omega$ (Fig. \ref{fig:w0=w}) occurs at $V_{dc}=0$ and corresponds to $0.82$ times vacuum fluctuations, i.e. $0.86$dB below vacuum. In all the measurements, the phase between the two microwave generators has been chosen to maximize the squeezing.

\emph{Theory.}
We have measured the amplitudes $A$ and $B$ of the two quadratures of the electromagnetic field generated by the tunnel junction, i.e. a property of the photon field. In order to link these with properties of the electrons crossing the junction, we suppose that the corresponding quantum operators $\hat{A}$ and $\hat{B}$ are related to the electron current operator at frequency $\omega$, $\hat{I}(\omega)$ by:
\begin{eqnarray}
\hat{A}=\frac{1}{\sqrt{2}} \left( \hat{I}(\omega)+\hat{I}^\dagger(\omega)\right) \nonumber\\
\hat{B}=\frac{i}{\sqrt{2}} \left( \hat{I}(\omega)-\hat{I}^\dagger(\omega)\right)
\label{eq:AB}
\end{eqnarray}%
with $\hat{A}^\dagger=\hat{A}$ and $\hat{B}^\dagger=\hat{B}$. The average commutator of those two observables  $\langle [ \hat{A},\hat{B}]\rangle = \langle[ \hat{I}(\omega),\hat{I}(-\omega)] \rangle=iS_{vac}(\omega)$ is non-zero, so the uncertainties in the measurement of $A$ and $B$ obey the Heisenberg uncertainty principle: $\Delta A^2\Delta B^2\geq S_{vac}^2$ with $\Delta O^2= \langle (\hat{O}-\langle \hat{O} \rangle)^2\rangle$ for $\hat{O}=\hat{A},\hat{B}$. These variances are related to current-current correlators:
\begin{eqnarray}
\Delta A^2 = \frac12\langle\{\hat I(\omega),\hat I(-\omega)\}\rangle + \frac12\left[\langle I(\omega)^2\rangle+\langle I(-\omega)^2\rangle\right] \nonumber \\
\Delta B^2 = \frac12\langle\{\hat I(\omega),\hat I(-\omega)\}\rangle - \frac12\left[\langle I(\omega)^2\rangle+\langle I(-\omega)^2\rangle\right]
\end{eqnarray}%
where the anti-commutator $\widetilde{S}=\langle\{\hat I(\omega),\hat I(-\omega)\}\rangle/2$ is the usual noise and $X=\left[ \langle I(\omega)^2\rangle+\langle I(-\omega)^2\rangle \right]/2$ the correlator describing the noise dynamics, studied in \cite{GR1,GR2}, which is non-zero only if $2\omega=p\omega_0$ with $p$ integer. In the absence of ac excitation, $X=0$ and $\Delta A^2=\Delta B^2=S$, which corresponds to $S_{vac}$ at $V_{dc}=0$. The condition for squeezing is thus $\widetilde{S}-X < S_{vac}$. $\widetilde{S}$ is given by Eq. (\ref{eq:Sp}) and $X$ is given by:
\begin{eqnarray}
X=&&\frac12 \sum_n J_n\left( \frac{e V_{ac}}{\hbar \omega_0} \right )J_{n+p}\left( \frac{e V_{ac}}{\hbar \omega_0} \right )
\nonumber\\
&&\left[ S_0\left(V_{dc}+\frac{\hbar}{e}(\omega+n\omega_0)\right)\right.
\nonumber\\
&&\left. +(-1)^{p} S_0\left(V_{dc}-\frac{\hbar}{e}(\omega+n\omega_0)\right) \right]
\label{eq:X}
\end{eqnarray}%
From Eqs.(\ref{eq:Sp}) and (\ref{eq:X}) one can calculate $\Delta A^2$ and $\Delta B^2$ as a function of $V_{dc}$ and $V_{ac}$. The theoretical predictions are shown as straight black lines together with the experimental data (colored symbols) on Figs. \ref{fig:w0=2w} ($\omega_0=2\omega$, i.e. $p=1$) and \ref{fig:w0=w} ($\omega_0=\omega$, i.e. $p=2$). In all the experimental conditions, theory and experiment agree very well. The existence of such squeezing has been predicted very recently \cite{BBRB}, and is a particular case of violation of a Cauchy-Schwarz inequality by electronic quantum noise. It is remarkable to note that the squeezing comes from the $X$ term which reflects the modulation of the shot noise by a time-dependent voltage. The origin of this term is in the shot noise itself, which comes from the granularity of the charge.

The optimal conditions for the observation of squeezing are very different for $p=1$ ($\omega_0=2\omega$, which corresponds to four-wave mixing) and $p=2$ ($\omega_0=\omega$, which corresponds to three-wave mixing), and can be easily understood at $T=0$, when $\widetilde{S}$ is a piecewise linear function. For $p=1$, $\widetilde{S}$ is independent of $V_{dc}$ as long as $V_{dc}<\hbar\omega$, while $X$ is a linear function of $V_{dc}$. As a result, the optimal $V_{dc}$ is $\hbar\omega/e$ (see Fig. \ref{fig:w0=2w}). We find that the maximal squeezing at $T=0$, $p=1$ corresponds to $\Delta A^2=0.62S_{vac}$, i.e. 2.09dB below vacuum. This corresponds to four-wave mixing. For $p=2$ $\widetilde{S}$ is minimal at $V_{dc}=0$ and increases as $|V_{dc}|$ while $X$ is maximal at $V_{dc}=0$ and decreases as $-|V_{dc}|$. Thus the optimal squeezing occurs at $V_{dc}=0$. We find that the maximal squeezing at $T=0$, $p=2$ corresponds to $\Delta A^2=0.73S_{vac}$, i.e. 1.37dB below vacuum. These results are independent of $\omega$ at zero temperature. Better squeezing can probably be achieved by exciting the sample with a periodic function that contains several harmonics \cite{GR_shaping}. Another possibility is to use another kind of sample, which would produce larger shot noise, thus increased non-linearity. To this respect, the tunnel junction is probably optimal in the category of normal conductors, but other samples with effective charge greater than one might be interesting to study.

\emph{Conclusion.}
The current/voltage fluctuations in the microwave domain generated by a conductor can be seen as a randomly fluctuating electromagnetic field. This is similar to the light emitted by the filament of a light bulb, though at a much lower frequency. We have shown that the microwave radiation generated by a quantum conductor (a dc+ac biased tunnel junction between normal metals placed at very low temperature) is squeezed, i.e. that the fluctuations on one quadrature can go below that of vacuum. Whereas any classical current (described a number) in a conductor generates a classical state of light \cite{Glauber}, we have demonstrated that a quantum current (described by an operator) can emit non-classical light. This offers a novel approach -based on the quantization of the electron charge, at the origin of shot noise- to the generation of quantum states of radiation, which are key for the development of quantum computation and communication.

We are very grateful to L. Spietz for giving us the tunnel junction (fabricated at NIST, Boulder, CO) used in the experiment. We acknowledge fruitful discussions with A. Bednorz, W. Belzig, A. Ferris and J. Gabelli and N. Godbout. This work was supported by the Canada Excellence Research Chairs program, the NSERC, the MDEIE, the INTRIQ of FRQNT and the Canada Foundation for Innovation.

%\vspace{-2mm}

\end{document}